\documentclass[preprintnumbers,aps,prb,preprint]{revtex4}

\usepackage{graphicx}
\usepackage{amsmath}
\usepackage{amssymb}

\begin{document}

\title{Stability of an oscillating tip in Non-Contact Atomic Force Microscopy~: theoretical and numerical investigations.}
\author{G. Couturier$^1$, L. Nony$^{2,\ast}$, R. Boisgard$^1$, J.-P. Aim\'{e}$^{1}$\\
 \small{$^1$ CPMOH, UMR CNRS 5798, Universit\'{e} Bordeaux I\\
 351, cours de la Lib\'{e}ration, 33405 Talence Cedex, FRANCE}\\
 \small{$^2$ L2MP, UMR CNRS 6137,
 Universit\'{e} d'Aix-Marseille III \\
 Facult\'{e} des Sciences de Saint-J\'{e}r\^{o}me, 13397 Marseille Cedex 20, FRANCE}\\
 \small{$^\ast$ To whom correspondence should be addressed; E-mail:
 laurent.nony@l2mp.fr}\\
 \textbf{published in Journal of Applied Physics 91(4), pp2537-2543 (2002)}}

\begin{abstract}
This paper is a theoretical and a numerical investigation of the stability
of a tip-cantilever system used in Non-Contact Atomic Force Microscopy
(NC-AFM) when it oscillates close to a surface. No additional dissipative
force is considered. The theoretical approach is based on a variationnal
method exploiting a coarse grained operation that gives the temporal
dependence of the nonlinear coupled equations of motion in amplitude and
phase of the oscillator. Stability criterions for the resonance peak are
deduced and predict a stable behavior of the oscillator in the vicinity of
the resonance. The numerical approach is based on results obtained with a
virtual NC-AFM developped in our group. The effect of the size of the stable
domain in phase is investigated. These results are in particularly good
agreement with the theoretical predictions. Also they show the influence of
the phase shifter in the feedback loop and the way it can affect the damping
signal.

\textbf{keywords} : NC-AFM, Variational principle, Stability, Virtual
machine, Phase shifter, Damping variations.
\end{abstract}

\maketitle

\section{Introduction}
In recent years, the use of the Non-Contact Atomic Force Microscopy (NC-AFM)
mode has shown that contrasts at the atomic scale could be achieved on
semiconductors and insulators surfaces \cite
{Geissibl95,Sugarawa95,Kitamura96,Bammerlin96,Schwarz99}. Experimental\ and
theoretical features dedicated to the description of this dynamic mode have
been widely discussed in previous papers \cite
{Albrecht91,Anczycowsky96,Geissib97,Wang97,Aime3_99,Sasaki99,Durig99}. In
particular, it was shown that the high sensitivity of the oscillating
tip-cantilever system (OTCS) was based on the value of the quality factor
and on its nonlinear dynamics in the vicinity of the surface \cite
{Aime2_99,Aime3_99}. Current considerations of the authors focus on the
origin of the increase of the damping signal when the tip comes close to the
surface, in the range of a few angstroms. Some claim that the origin of this
apparent increase could be due to the hysteretic behavior of the OTCS \cite
{Aime3_99,Gauthier00}. These interpretations implicitly rise the question of
the stability of the OTCS when it is at the proximity of the surface.

The aim of this paper is to show from a theoretical and a numerical point of
view that the nonlinear dynamics of the OTCS leads to various stability
domains of its resonance peak that may help to understand the reason why the
NC-AFM mode, while being so sensitive, keeps, in most of cases, a stable
behavior. In other words, this work is an attempt to show that, if no
additional dissipative force is considered between the tip and the surface,
an apparent increase of the damping signal cannot be the consequence of the
nonlinear behavior of the OTCS.

\bigskip

The paper is organized as follow. The first part is dedicated to a
description of the nonlinear behavior of the OTCS at the proximity of the
surface. To do so, a specific theoretical frame based on a variational
method using a coarse grained operation has been developed. This gives the
explicit temporal dependance of the OTCS equations of motion. These
equations are the basis to analyze the stability of the stationary state. A
large part of this work is detailed in ref.\cite{Nony2_01}. Experimentally,
as the NC-AFM mode requires the use of a feedback loop to maintain constant
the amplitude at the resonance frequency, which in turn requires to maintain
constant the phase of the oscillator around its resonance value, e.g. $-\pi
/2~$rd, the phase variations of the OTCS will be extracted and discussed.
The second part of the paper deals with numerical results obtained with the
virtual NC-AFM \cite{Couturier01} which is very similar to the experimental
machine. These results show unambiguously the contribution of the phase
shifter in the feedback loop and the way it can lead to damping
variations.\newpage

\section{Theoretical approach of the NC-AFM}

The present section is divided into three parts. The first one details the
specific theoretical frame for the obtention of the equations of motion in
amplitude and phase of the OTCS. A coarse-grained method gives the equation
describing the time evolution of the stationary state of the OTCS as a
function of the coupling term between the tip and the surface. The second
part is a description of the distortion of the resonance peak as a function
of the distance. This part was detailed in ref.\cite{Nony2_01} so that only
the main results are given. The results provide the basis of the discussion
about the stability of the branches which is detailed in the third part.

\subsection{Theoretical frame}

We search a solution to the temporal evolution of the OTCS by using a
variational solution based on the principle of least action. Even though
this approach exploits the same physical concepts than the one which had led
to the coupled equations in amplitude and phase of the stationary state of
the OTCS \cite{Aime3_99,Nony99}, it appears to be more general since here,
the temporal dependance is explicitly obtained. We start from the definition
of the action of the OTCS coupled to an interaction potential~:
\begin{equation}
S=\int_{t_{a}}^{t_{b}}\mathcal{L}\left( z,\dot{z},t\right) dt\text{,}
\end{equation}
where $\mathcal{L}$ is the Lagrangian of the system and $z\left( t\right) $
the position of the tip with time \cite{Nony99}~:

\begin{eqnarray}
\mathcal{L}\left( z,\dot{z},t\right) &=&\mathcal{T}-\mathcal{V}+\mathcal{W}
\label{equLagrangien} \\
&=&\frac{1}{2}m^{\ast }\dot{z}\left( t\right) ^{2}-\left[ \frac{1}{2}%
k_{c}z\left( t\right) ^{2}-z\left( t\right) \mathcal{F}_{exc}\cos \left(
\omega t\right) +V_{int}\left[ z\left( t\right) \right] \right] -\frac{%
m^{\ast }\omega _{0}}{Q}z(t)\underline{\dot{z}}\left( t\right)  \nonumber
\end{eqnarray}
$\omega _{0}$, $Q$, $m^{\ast }$ and $k_{c}=m^{\ast }\omega _{0}^{2}$ are
respectively the resonance pulsation, quality factor, effective mass and
cantilever' stiffness of the OTCS. $\mathcal{F}_{exc}$ and $\omega $ are the
external drive force and drive pulsation. Due to the large quality factor,
we assume that a typical temporal solution is on the form~:
\begin{equation}
z\left( t\right) =A\left( t\right) \cos \left[ \omega t+\varphi \left(
t\right) \right] \text{,}  \label{equfonctionharmonique}
\end{equation}
where $A\left( t\right) $ and $\varphi \left( t\right) $ are assumed to be
slowly varying functions with time compared to the period $T=2\pi /\omega $.
The underlined variables of $\underline{\dot{z}}\left( t\right) $ in equ.\ref
{equLagrangien}, e.g. $\underline{A}\left( t\right) $, $\underline{\varphi }%
\left( t\right) $, $\underline{\dot{A}}\left( t\right) $ and $\underline{%
\dot{\varphi}}\left( t\right) $, are calculated along the physical path,
thus they are not varied into the calculations \cite{Goldstein80}.

To describe the interaction between the tip and the surface, the attractive
coupling force is assumed to derive from a sphere-plane interaction
involving the disperse part of the Van der Waals potential \cite
{Israelachvili92}~:

\begin{equation}
V_{int}\left[ z\left( t\right) \right] =-\frac{HR}{6\left[ D-z\left(
t\right) \right] }  \label{equpotVdWdisp}
\end{equation}
$H$, $R$ and $D$ are the Hamaker constant, the tip's apex radius and the
distance between the surface and the equilibrium position of the OTCS.

The equations of motion in amplitude and phase of the OTCS are obtained by
considering the following coarse-grained operation. Let's assume a long
duration $\Delta t=t_{b}-t_{a}$ with $\Delta t\gg T$ and calculate the
action as a sum of small pieces of duration $T$~:
\begin{equation}
S=\sum_{n}\int_{nT}^{\left( n+1\right) T}\mathcal{L}\left( z,\dot{z}%
,t\right) dt=\sum_{n}\left( \frac{1}{T}\int_{nT}^{\left( n+1\right) T}%
\mathcal{L}\left( z,\dot{z},t\right) dt\right) T=\sum_{n}\mathcal{L}_{e}T
\end{equation}
$\mathcal{L}_{e}$ is the mean Lagrangian during one period and\ appears as
an effective Lagrangian for a large time scale compared to the period. Owing
to the quasi-stationary behavior of the amplitude and the phase over the
period, the effective Lagrangian is calculated by keeping them constant
during the integration. The calculations give~:
\begin{eqnarray}
\mathcal{L}_{e}\left( A,\dot{A},\varphi ,\dot{\varphi}\right) &=&\frac{%
m^{\ast }}{4}\left[ \dot{A}^{2}+A^{2}\left( \omega +\dot{\varphi}^{2}\right) %
\right] -\frac{k_{c}A^{2}}{4}+\frac{\mathcal{F}_{exc}A\cos \left( \varphi
\right) }{2}-\frac{1}{T}\int_{0}^{T}V_{int}\left[ z\left( t\right) \right] dt
\nonumber \\
&&-\frac{m^{\ast }\omega _{0}}{2Q}\left[ A\underline{\dot{A}}\cos \left(
\varphi -\underline{\varphi }\right) -A\underline{A}\left( \omega +%
\underline{\dot{\varphi}}\right) \sin \left( \underline{\varphi }-\varphi
\right) \right]
\end{eqnarray}
Note that the effective Lagrangian is now a function of the new generalized
coordinates $A$, $\varphi $ and their associated generalized velocities $%
\dot{A}$, $\dot{\varphi}$. At this point, remembering that the period is
small regardless to $\Delta t=t_{b}-t_{a}$ during which the total action is
evaluated, the continuous expression of the action is~:
\begin{equation}
S=\int_{t_{a}}^{t_{b}}\mathcal{L}_{e}\left( A,\dot{A},\varphi ,\dot{\varphi}%
\right) d\tau \text{,}
\end{equation}
where the measure $d\tau $ is such that $T\ll d\tau \ll \Delta t$.

Applying the principle of least action $\delta S=0$ to the functional $%
\mathcal{L}_{e}$, we obtain the Euler-Lagrange equations for the effective
Lagrangian. Thus, the amplitude and phase equations of motion of the OTCS
coupled to the surface through an interaction involving the disperse part of
the Van der Waals potential are obtained~:

\begin{equation}
\left\{
\begin{array}{c}
\ddot{a}=\left[ \left( u+\dot{\varphi}\right) ^{2}-1\right] a-\dfrac{\dot{a}%
}{Q}+\dfrac{\cos \left( \varphi \right) }{Q}+\dfrac{a\kappa _{a}}{3\left(
d^{2}-a^{2}\right) ^{3/2}} \\
\ddot{\varphi}=-\left( \dfrac{2\dot{a}}{a}+\dfrac{1}{Q}\right) \left( u+\dot{%
\varphi}\right) -\dfrac{\sin \left( \varphi \right) }{aQ}
\end{array}
\right. \text{,}  \label{equequationmouvementadim}
\end{equation}
In equ.\ref{equequationmouvementadim}, specific notations were used. $%
d=D/A_{0}$ is the reduced distance between the location of the surface and
the equilibrium position of the OTCS normalized to the resonance amplitude $%
A_{0}=Q\mathcal{F}_{exc}/k_{c}$, $a=A/A_{0}$ is the reduced amplitude, $%
u=\omega /\omega _{0}$ is the reduced drive frequency normalized to the
resonance frequency of the free OTCS and $\kappa _{a}=HR/\left(
k_{c}A_{0}^{3}\right) $ is the dimensionless parameter that characterizes
the strength of the interaction.

\subsection{Resonance frequency shift\label{subsectionresonancepeak}}

The equations of motion of the stationary solutions $a$ and $\varphi $ are
obtained by setting $\dot{a}=\dot{\varphi}=0$ and $\ddot{a}=\ddot{\varphi}=0$
in equ.\ref{equequationmouvementadim} and lead to two coupled equations of
the sine and cosine of the phase of the OTCS previously calculated \cite
{Nony99}~:

\begin{equation}
\left\{
\begin{array}{l}
\cos \left( \varphi \right) =Qa(1-u^{2})-\dfrac{aQ\kappa _{a}}{3\left(
d^{2}-a^{2}\right) ^{3/2}} \\
\sin \left( \varphi \right) =-ua
\end{array}
\right. \text{,}  \label{equcossin}
\end{equation}
Solving equ.\ref{equcossin} gives the relationship between the frequency and
the amplitude at a given distance $d$ \cite{Aime3_99}~:

\begin{equation}
u_{\pm }\left( a\right) =\sqrt{\frac{1}{a^{2}}-\frac{1}{4Q^{2}}\left( 1\mp
\sqrt{1-4Q^{2}\left( 1-\frac{1}{a^{2}}-\frac{\kappa _{a}}{3\left(
d^{2}-a^{2}\right) ^{3/2}}\right) }\right) ^{2}}  \label{equuapprochevar}
\end{equation}
The signs plus and minus are deduced from the sign of $\cos \left( \varphi
\right) $ and correspond to values of the phase ranging from $0$ to $-90%
{{}^\circ}%
$ ($u_{-}$, $\cos \left( \varphi \right) >0$) or from $-90%
{{}^\circ}%
$ to $-180%
{{}^\circ}%
$ ($u_{+}$, $\cos \left( \varphi \right) <0$), in agreement with the sign
convention of the phase in equ.\ref{equfonctionharmonique}. From equ.\ref
{equuapprochevar} is calculated the resonance peak at any reduced distance
for a given strength of the sphere-surface interaction and equ.\ref
{equcossin} give the associated phase variations. The two branches define
the distortion of the resonance peak as a function of $d$. $u_{-}$ gives the
evolution of the resonance peak for frequency values below the resonance one
and $u_{+}$ for frequency values above the resonance.

\begin{figure}[h]
\includegraphics[width=10cm]{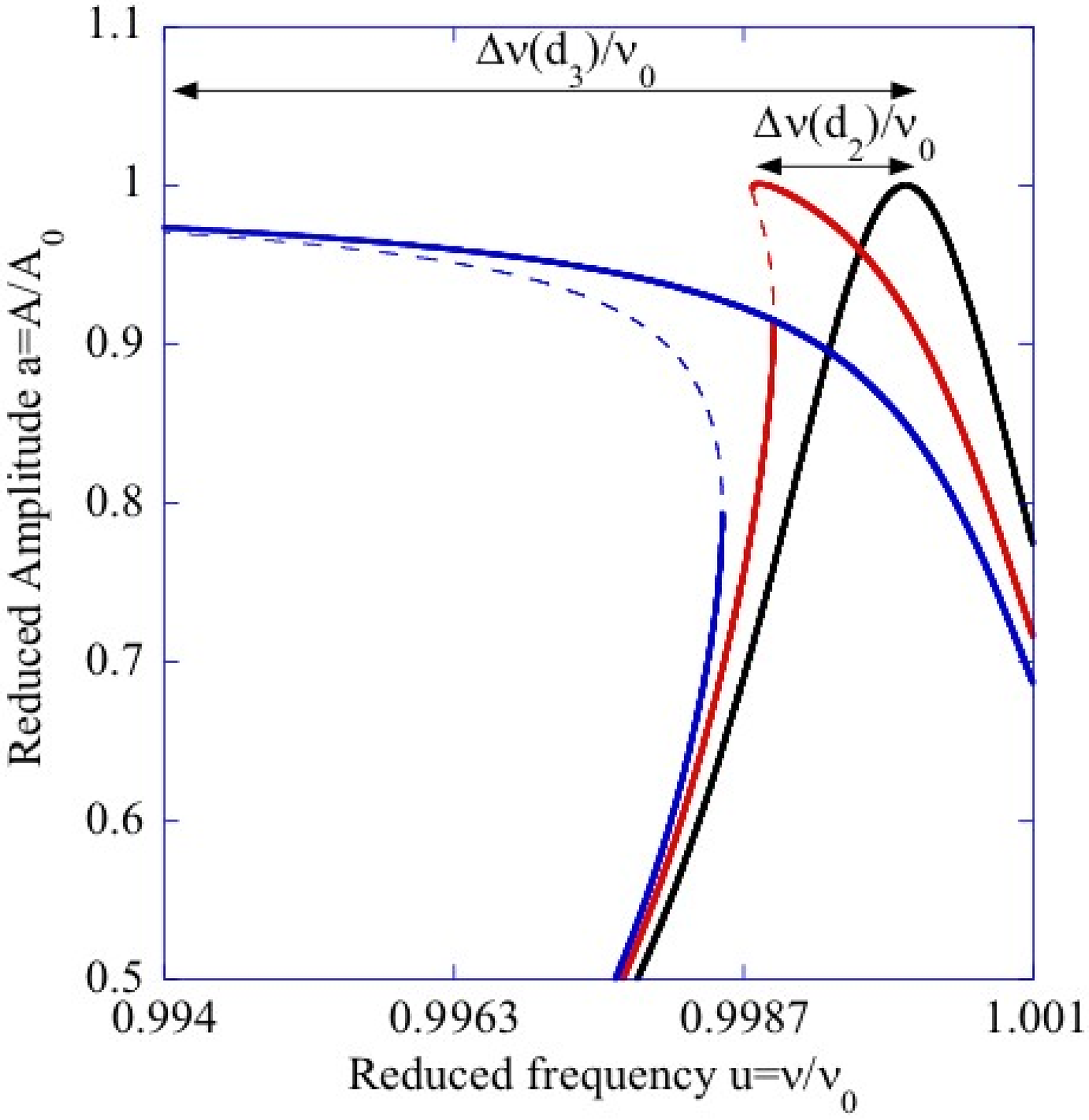}\\
  \caption{Distortion of the resonance peak computed from equ.\ref
{equuapprochevar} for three values of the distance, $d_{1}=2$, $d_{2}=1.11$
and$\ d_{3}=1.012$. The numerical parameters are $A_{0}=20nm$, $Q=400$ and $%
\kappa _{a}=8.10^{-4}$. For an attractive coupling, the peak is more and more distorted towards the low frequencies as $d$ is reduced, e.g. the
surface is approached. For each value of $d$, the unstable domains of $u_{-}$ are shown with dashed lines. The arrows indicate the resonance
frequency shift vs. $d$.}\label{figreso}
\end{figure}

In fig.\ref{figreso} is given the distortion of the resonance peak vs. the
reduced distance $d$. For large values of $d$, e.g. when the surface is far
from the OTCS, the nonlinear effects are negligible and the peak keeps a
well-defined Lorentzian shape (see equ.\ref{equuapprochevar} with $\kappa
_{a}=0$). When the OTCS is approached towards the surface, because the
interaction is attractive, the resonance peak starts to distort towards the
low frequencies. The distortion of the peak increases as $d$ decreases. On
the fig.\ref{figreso}, the branches that are supposed unstable are shown
with dashed lines.

Using equ.\ref{equuapprochevar},\ the resonance frequency shift as a
function of the distance $d$ is obtained by setting $a=1$. This former
condition ensures the required condition for the NC-AFM mode. Thus, the
normalized frequency shift, $\left( \nu -\nu _{0}\right) /\nu _{0}$,\ is
given by $u-1$ \cite{Aime3_99}~:

\begin{equation}
u_{_{\pm }}\left( d\right) -1=\sqrt{1-\frac{1}{4Q^{2}}\left( 1\mp \sqrt{1+%
\frac{4}{3}\frac{Q^{2}\kappa _{a}}{\left( d^{2}-1\right) ^{3/2}}}\right) ^{2}%
}-1  \label{equshiftattractif}
\end{equation}
The frequency shift given by equ.\ref{equshiftattractif} vs. $d$ can be
deduced from the fig.\ref{figreso} (see the arrows on the figure). Following
the previous discussion about the stability of the different parts of the
resonance peak during the distortion, since the measure is performed as a
function of $d$ with $a=1$, no bistable behavior can be observed.

However, in the vicinity of the resonance the branches $u_{+}$ and $u_{-}$
become very close (see for instance fig.\ref{figreso} with $d_{3}=1.012$).
Therefore, even with an oscillation amplitude kept constant, question rises
about the ability of the OTCS to remain on the same branch. Qualitatively,
one may expect that around $a\cong 1$,\ the branch $u_{-}$ is unstable and $%
u_{+}$ is stable (see fig.\ref{figreso}). If this is true, any small
fluctuation of the oscillation amplitude might produce a jump from one
branch to the other one as discussed in refs.\cite{Aime3_99,Gauthier00}.
Since the branch $u_{-}$ seems to be unstable, a jump to this branch should
lead to an abrupt decrease of the amplitude, which in turn might produce an
apparent abrupt increase of the damping signal as a consequence of the
hysteretic behavior. Because such a jump is, in most of cases, never
observed, it becomes useful to determine more accurately the stability of
the two branches.

\subsection{Stability criterions\label{sectionstability}}

\begin{figure}[h]
\includegraphics[width=10cm]{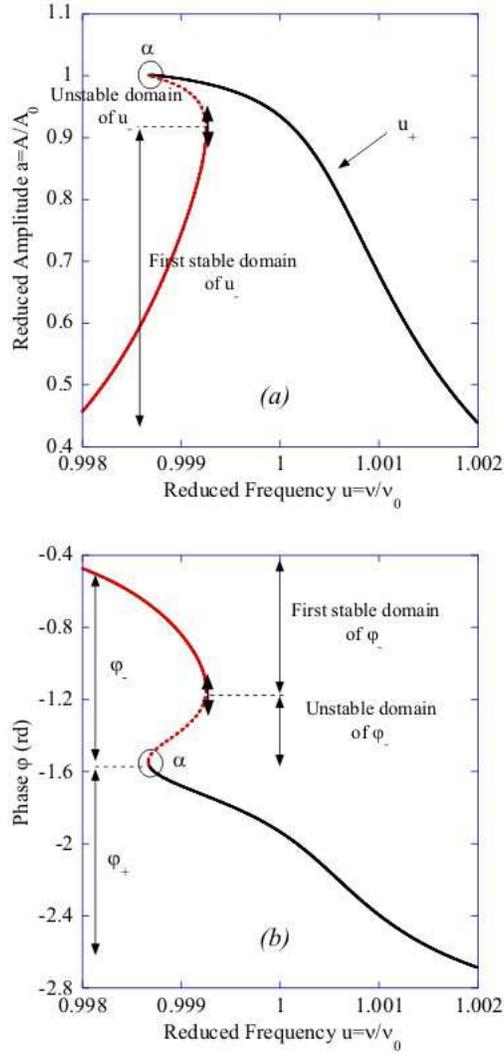}\\
  \caption{\textit{(a)}- Distortion of the resonance peak computed from equ.%
\ref{equuapprochevar}. The numerical parameters are $d=1.05$, $A_{0}=10nm$, $%
Q=500$ and $\kappa _{a}=2.5.10^{-4}$. The stability criterions foresee that $%
u_{+}$ is always stable whereas $u_{-}$ exhibits two stable domains (continuous lines) and one unstable (dashed lines). The domains are
separated by the spots where the derivative $da/du_{-}$ is infinite. \textit{%
(b)}- Distortion of the phase curve computed from equs.\ref{equcossin}
associated to the resonance peak. As a consequence of the stability of $%
u_{+} $, $\varphi _{+}$ is always stable whereas $\varphi _{-}$ exhibits two stable domains and one unstable.}\label{figstabilitebranches}
\end{figure}

\begin{figure}[h]
\includegraphics[width=10cm]{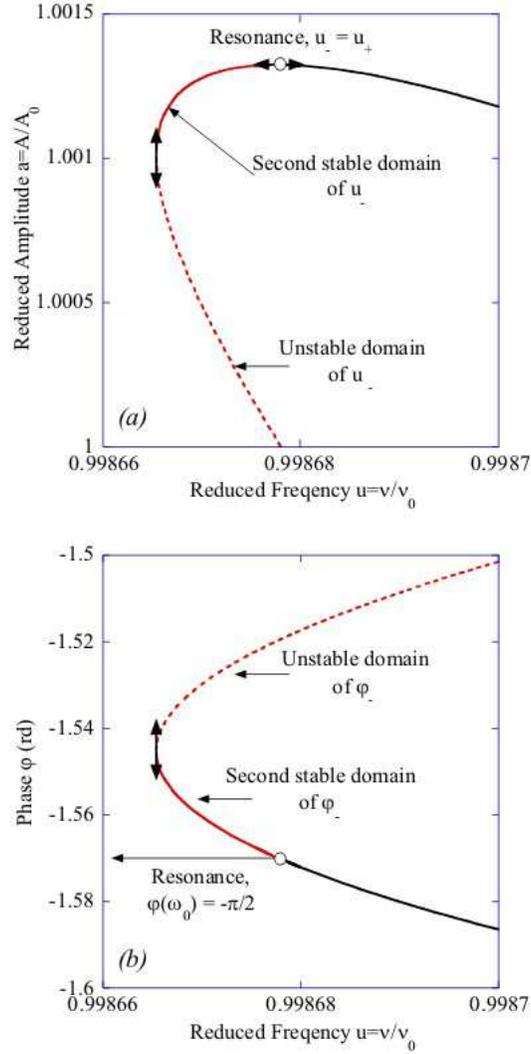}\\
  \caption{\textit{(a)}- Zoom in the region $\alpha $ of the resonance peak. As $da/du_{-}$ becomes infinite again, the criterions define a new domain of
$u_{-}$ which is stable. The resonance is located where $u_{+}=u_{-}$. \textit{(b)}- Zoom in the region $\alpha $ of the phase curve. The
resonance is located at $-\pi /2$ where $\varphi _{+}=\varphi _{-}$ and belongs to a stable domain.}\label{figstabilitebranchesz2}
\end{figure}

The stability of the branches $u_{\pm }$ of the resonance peak is obtained
from equations of motion of the OTCS (see equ.\ref{equequationmouvementadim}%
). By linearizing these equations around the stationary solution (now
identified by the index ``$s$'') and using classical considerations of the
linear theory, one gets the stability criterions of the branches $u_{\pm }$
. The stability criterions can be expressed from the derivatives $%
da_{s}/du_{\pm }$ of the branches and reduced to the simple expression~\cite
{Nony2_01}:

\begin{equation}
\left\{
\begin{array}{c}
\dfrac{da_{s}}{du}>0\qquad \text{and}\qquad \cos \left( \varphi _{s}\right)
>a_{s}/\left( 2Q\right) \qquad (i) \\
\text{or} \\
\dfrac{da_{s}}{du}<0\qquad \text{and}\qquad \cos \left( \varphi _{s}\right)
<a_{s}/\left( 2Q\right) \qquad (ii)
\end{array}
\right.  \label{equconditionstabilite2}
\end{equation}
The figs.\ref{figstabilitebranches}\textit{(a)}\ and \textit{(b)} show the
distortion of the resonance peak and of the associated phase curve,
respectively. The figs.\ref{figstabilitebranchesz2}\textit{(a) }and\textit{\
}\ref{figstabilitebranchesz2}\textit{(b) }are zooms on the region $\alpha $
of the figs.\ref{figstabilitebranches}\textit{(a)}\ and \textit{(b)},
respectively.

For the branch $u_{+}$, $da_{s}/du_{+}$ being always negative and the
associated value of the phase being always defined beyond $-\pi /2$ (see
section \ref{subsectionresonancepeak}), the criterion $(ii)$ implies that $%
u_{+}$ is always stable, whatever the value of $a_{s}$.

For $u_{-}$, the sign of the derivative changes twice. For this branch, the
phase is always defined above $-\pi /2$. Therefore on the lower part of the
branch (small $a$), $da_{s}/du_{-}>0$ and the criterion $(i)$ indicates that
the branch is locally stable. When $da_{s}/du_{-}$ becomes negative (see fig.%
\ref{figstabilitebranches}\textit{(a)}), because $\varphi _{s}>-\pi /2$ (see
fig.\ref{figstabilitebranches}\textit{(b)}), the criterion $(i)$ is no more
filled. As a consequence, $u_{-}$ is locally unstable and the instability is
precisely located where the infinite tangent appears. On the upper part of
the resonance peak, the curvature of $u_{-}$ changes again and $%
da_{s}/du_{-}>0$ (see fig.\ref{figstabilitebranchesz2}\textit{(a)}),
implying that it is again a locally stable domain. Thus the branches $u_{-}$
and $\varphi _{-}$ exhibit two stable domains and one unstable.

Note also that the resonance condition is deduced from $da_{s}/du=0$ which
implies $\cos \left( \varphi _{s}\right) =a_{s}/\left( 2Q\right) $, or
equivalently $u_{-}=u_{+}$ or again $\varphi _{-}=\varphi +$. This equality
is the usual resonance condition of a free harmonic oscillator. If $a_{s}=1$%
, e.g. without any coupling, the resonance phase is therefore $\varphi
_{s}=\arccos \left[ 1/\left( 2Q\right) \right] $. For the OTCS we used, $%
Q\simeq 500$, and so $\varphi _{s}\cong -\pi /2$. But taking into account
the fact that the coupling only slightly modifies the value of the resonance
amplitude, $a_{s}\simeq 1.0013$ (see fig.\ref{figstabilitebranches}\textit{%
(a)}), we still obtain $\varphi _{s}\cong -\pi /2$ so that we can consider
that the nonlinear resonance is always given by the relationship $\varphi
_{s}=-\pi /2$.

\bigskip

Therefore the theoretical approach foresees that $u_{+}$ is always stable
but that also a small domain of $u_{-}$ around the resonance value remains
stable. If the resonance value would have been located at the point where $%
da_{s}/du_{-}$ is infinite, an infinitely small fluctuation would have been
able to generate an abrupt increase of the damping signal as discussed
previously and suggested in ref.\cite{Aime3_99}, or more recently in ref.
\cite{Gauthier00}. Experimentally, an electronic feedback loop keeps
constant the amplitude of the OTCS so that its phase is located around $-\pi
/2$ (see section below). As a consequence, question rises about the size of
the stable domain in phase around $-\pi /2$. If any fluctuation around $-\pi
/2$ makes the phase going beyond the stable domain, the OTCS behavior
becomes unstable. For $Q=500$, the size of the stable domain is of about $%
2.6.10^{-2}~$rd (see fig.\ref{figstabilitebranchesz2}\textit{(b)}) whereas
it's reduced to $2.6.10^{-3}~$rd for $Q=5000$ (data not shown). Thus, if the
electronic loop is able to control the phase locking with a better accuracy
than $2.6.10^{-3}~$rd, the OTCS will be locked on a stable domain.

Therefore, if the setpoint of the oscillator is properly located at the $%
-\pi /2$ value throughout an experiment, this value corresponds to a stable
domain and consequently will neither give rise to amplitude or phase
variations nor to damping variations.

\section{Virtual NC-AFM results\label{sectionVirtual}}

In a recent paper, we have described a virtual NC-AFM machine built using
the \textit{Matlab} language and the \textit{Simulink} toolbox \cite
{Couturier01}. This machine is very similar to our own experimental hybrid
machine built with \textit{Digital Instruments }\cite{Digital}\textit{\ }and
\textit{Omicron} \cite{OmicronEnglish}\ blocks. The virtual machine has been
extensively used to study the frequency shift and the damping signal in the
approach-retract mode. Two types of situations have been investigated~: $i-$
the first one corresponds to the case where no dissipative force is
introduced in the tip-surface interaction, $ii-$ the second one deals with
dissipative forces. In both cases an attractive sphere-plane Van der Waals
interaction is taken into account.

In spite of previous results that have already shown that the damping signal
could be considered as a constant when no dissipative force was introduced
\cite{Couturier01}, here we want to investigate with the virtual machine the
stability of the OTCS by looking accurately at its phase variations within
the electronic feedback loop that maintains constant the amplitude of the
oscillations and compare these results with the theoretical predictions. To
do so, we still do not consider any additional dissipative force.

\bigskip

The theoretical results have led to the conclusion that, provided that the
OTCS phase is in the close vicinity of $-\pi /2$, this setpoint corresponds
to a stable branch. As a consequence, the questions are~: $i-$ What is the
part of the feedback loop that controls the size of the phase stable domain
of the OTCS around -$\pi /2~$rd ? and $ii-$ Would it be possible to change
the parameters of this element in order to change the size of the phase
stable domain and thus observe phase variations? As a consequence of these
changes, variations of the damping signal should also be observed.

\subsection{The phase shifter of the feedback loop}

\begin{figure}[h]
\includegraphics[width=10cm]{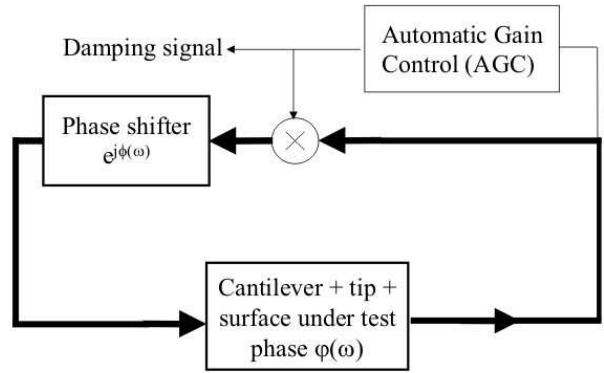}\\
  \caption{Schematic diagram of the feedback loop used in the virtual NC-AFM
which is very similar to the one of the experimental machine.}\label{figblanck1}
\end{figure}

In fig.\ref{figblanck1} is drawn a very simplified schematic diagram of the
feedback loop of the NC-AFM (for more details, see ref.\cite{Couturier01}).
Usually, the phase $\phi \left( \omega \right) $ of the phase shifter
transfer function is adjusted to $-3\pi /2$ so that the loop oscillates at $%
\nu _{0}$ which is the free resonance frequency of the cantilever,
corresponding to a tip-surface distance $D\rightarrow \infty $. We recall
that the oscillations of the loop are ruled by the relation~:

\begin{equation}
\phi \left( \omega \right) +\Bbb{\varphi }\left( \omega \right) =0\pm 2n\pi
\text{,}  \label{equConditionReso}
\end{equation}
where $n$ is an integer and $\Bbb{\varphi }\left( \omega \right) $ is the
phase difference between the oscillations and the excitation of the
cantilever. If the setpoint is fixed to the resonance frequency, then $\Bbb{%
\varphi }\left( \omega _{0}\right) =-\pi /2$. The phase adjustment in the
\textit{Omicron} electronics is obtained by changing the bias of varicap
diodes \cite{OmicronDoc}. The phase shifter transfer function in terms of
the $p$ Laplace variable can be written as $H\left( p\right) =\left( \dfrac{%
1-\tau p}{1+\tau p}\right) ^{2}$, the time constant $\tau $\ being adjusted
by the user such that, at the resonance~:

\begin{equation}
\phi \left( \omega _{0}\right) =-4\arctan \left( \tau \omega _{0}\right) =-%
\dfrac{3\pi }{2}  \label{equConditionPhi}
\end{equation}

When the tip-surface distance $D$ is reduced, due to the coupling, $\omega
_{0}$ decreases. As a consequence, $\phi \left( \omega _{0}\right) $ and $%
\Bbb{\varphi }\left( \omega _{0}\right) $ and are no more equal to $-3\pi /2$
and $-\pi /2$ respectively. According to equ.\ref{equConditionReso}, the
variation of $\Bbb{\varphi }\left( \omega \right) $ is governed by the one
of $\phi \left( \omega \right) $. Assuming a small variation around the
resonance frequency $\Delta \omega =\omega -\omega _{0}$, one gets~:

\begin{equation}
\phi \left( \omega \right) \simeq -3\pi /2-\dfrac{4\tau }{1+\left( \tau
\omega _{0}\right) ^{2}}\Delta \omega
\end{equation}
As $D$ decreases, $\Delta \omega $ is negative. Therefore $\phi \left(
\omega \right) $ becomes larger than $-3\pi /2$ and $\Bbb{\varphi }\left(
\omega \right) $ smaller than $-\pi /2$. The decrease of $\Bbb{\varphi }%
\left( \omega \right) $, $\Bbb{\varphi }\left( \omega \right) \lesssim -\pi
/2$, means that the phase of the OTCS follows the phase branch associated to
$u_{+}$, $\varphi _{+}$, which is always stable (see fig.\ref
{figstabilitebranches}\textit{(b)}). Thus the loop is always stable.
Moreover, the hypothesis implying that $\Bbb{\varphi }\left( \omega \right) $
keeps a value close to $-\pi /2$ is a very good assumption. To proof that,
let us consider for instance $\nu _{0}=150$~kHz, which is a reasonable value
for a cantilever. Therefore $\tau =2.56.10^{-6}$~s (see equ.\ref
{equConditionPhi}). Assuming now a large frequency shift, $\Delta \nu =-200$%
~Hz, we get $\Delta \phi \left( \omega \right) =1.9.10^{-3}~$rd and
therefore $\Delta \Bbb{\varphi }\left( \omega \right) =-1.9.10^{-3}~$rd. In
spite of the rough assumption of a first order expansion of the phase of the
phase shifter, the typical phase variations of the OTCS around the nonlinear
resonance are less than $2.10^{-3}~$rd. This implies that the machine
properly follows the nonlinear resonance, even when large frequency shifts
are considered.

\begin{figure}[h]
\includegraphics[width=10cm]{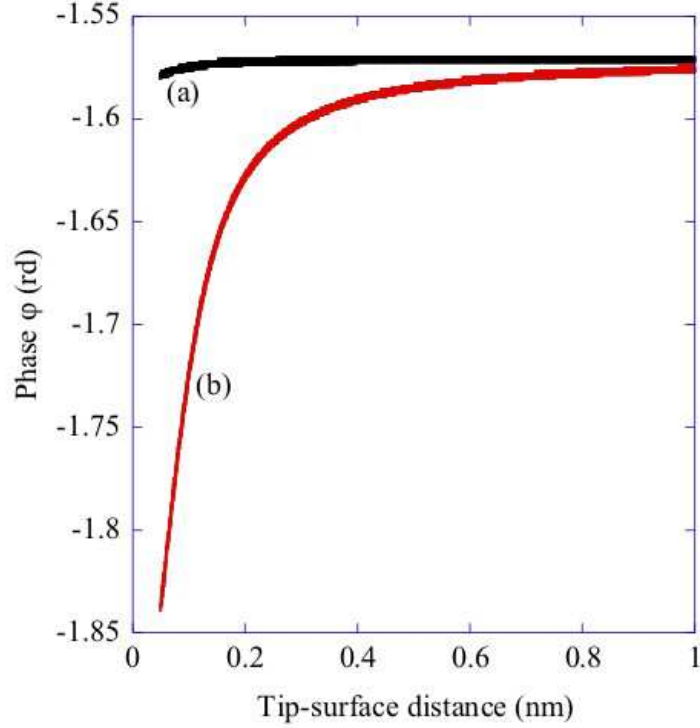}\\
  \caption{Variations of the phase of the OTCS $\varphi \left( \omega
\right) $ within the feedback loop vs. the distance $D$ computed from the
virtual NC-AFM. The numerical parameters are~: resonance amplitude $%
A_{0}=15~ $nm, spring constant $k_{c}=40$~N.m$^{-1}$, quality factor $Q=5000$%
, tip's radius $R=10$~nm and Hamaker constant $H=2.10^{-19}$~J. Curve [a]~:
The phase $\phi \left( \omega \right) $ of the transfer function $H\left(
p\right) $ of the phase shifter is the one given by equ.\ref{equConditionPhi}%
, e.g. is similar to the experimental machine. As $D$ decreases, $\varphi
\left( \omega \right) $ becomes weakly smaller than $-\pi /2$ (less than $%
2.10^{-3}~$rd), therefore follows the stable branch $\varphi _{+}$. The
machine follows accurately the setpoint, which is always stable, even when
the tip is in the very close vicinity of the surface. The associated damping
variation nearly no varies (see fig.\ref{figblanck3}). Curve [b]~: $\phi
\left( \omega \right) $ is the phase of $H\left( p\right) $ whose expression
is given by equ.\ref{equtransferfunction}. Around $-\pi /2$, the slope $%
d\phi \left( \omega \right) /d\omega $ is larger than in case [a] so that $%
\varphi \left( \omega \right) $ decreases more quickly. As a consequence, the damping signal increases.}\label{figblanck2}
\end{figure}

\begin{figure}[h]
\includegraphics[width=10cm]{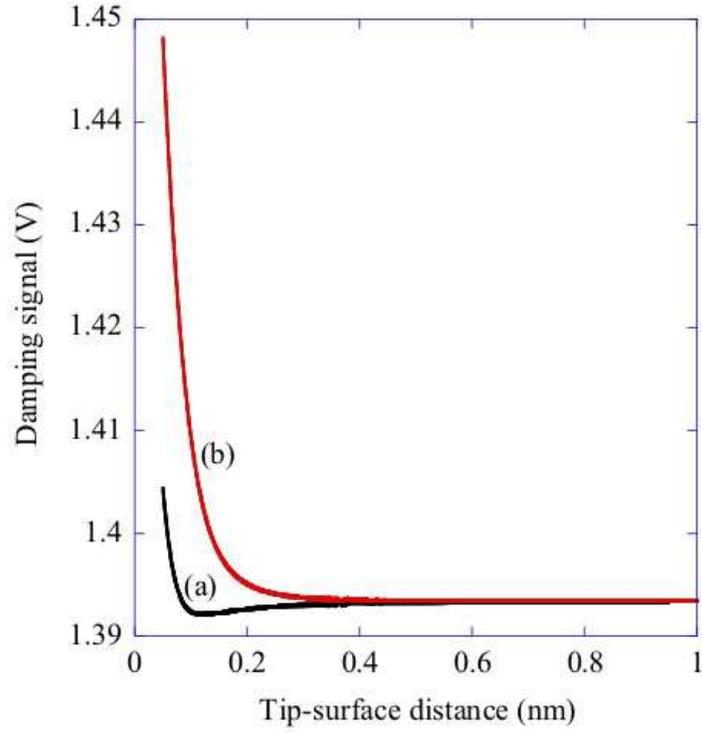}\\
  \caption{Variations of the damping signal vs.\thinspace the distance $D$. Curve [a]~: No damping variation is observed if the phase of the
virtual NC-AFM phase shifter is similar to the one of the experimental machine. In the very close vicinity of the surface, as $\varphi \left(
\omega _{0}\right) \gtrapprox -\pi /2$, e.g. the amplitude of the oscillations slightly decreases, a weak increase is observed. Curve [b]~: As
$\varphi \left( \omega _{0}\right) $ varies more quickly than in curve [a] due to the different expression of $H\left( p\right) $, a larger
increase of the damping is obtained.}\label{figblanck3}
\end{figure}

The curves [a] in figs.\ref{figblanck2} and \ref{figblanck3} show the phase $%
\Bbb{\varphi }\left( \omega \right) $ and the damping signal vs. the
distance $D$, respectively. As expected, the variations are very weak.

\subsection{"Controlled" damping variations}

If we want to observe the phase instability predicted by the theoretical
calculations, the phase shifter transfer function should have been on the
form $d\phi \left( \omega \right) /d\omega >0$ around $\omega _{0}$. A
possible expression of such a transfer function would be~: $H\left( p\right)
=\left( \dfrac{1+\tau p}{1-\tau p}\right) ^{2}$. Experimentally, this form
is not feasible and even if it were, the loop would become unstable and
therefore no stationary state could be reached. The reason is that the
inverse Laplace transform of $1/\left( 1-\tau p\right) $\ varies as $%
e^{t/\tau }$ which diverges as $t\rightarrow \infty $.

\bigskip

In the virtual machine, it is possible to implement a phase shifter with a
slope $d\phi \left( \omega \right) /d\omega $ larger than the phase shifter
built by \textit{Omicron}. We have retained the following transfer function
which is easy to do with electronic components~:

\begin{equation}
H\left( p\right) =\left( \dfrac{p^{2}-\dfrac{\omega _{1}}{Q_{1}}p+\omega
_{1}^{2}}{p^{2}+\dfrac{\omega _{1}}{Q_{1}}p+\omega _{1}^{2}}\right)
\label{equtransferfunction}
\end{equation}
The parameters $\omega _{1}$ and $Q_{1}$ may be adjusted to obtain for
instance $\phi \left( \omega _{0}\right) =-3\pi /2$. The phase of the
transfer function is then~:

\begin{equation}
\phi \left( \omega \right) =-2\arctan \left( \dfrac{\omega _{1}\omega }{%
Q_{1}\left( \omega _{1}^{2}-\omega ^{2}\right) }\right)
\end{equation}
For a small frequency shift, $d\phi \left( \omega \right) /d\omega \simeq
-4Q_{1}/\omega _{0}$. Keeping the same values than previously $\nu _{0}=150$%
~kHz and $\Delta \nu =-200$~Hz and assuming $Q_{1}=50$, we now obtain a
change $\Delta \phi \left( \omega \right) $ of about $0.26$~rd.
Consequently, the change $\Delta \Bbb{\varphi }\left( \omega \right) $
becomes larger (see the curve [b] in fig.\ref{figblanck2}) and we now
observe an increase of the damping signal as shown in fig.\ref{figblanck3},
curve [b].

The previous examples are pedagogical cases for which an arbitrary large
value of the slope of the phase of the phase shifter was considered. The
ideal phase shifter should maintain the phase $\phi \left( \omega
_{0}\right) $ at $-3\pi /2$ so that the frequency of the loop remains equal
to the resonance frequency of the cantilever. This is not possible in
practice, however it is clear that the solution retained by \textit{Omicron}
is very close to the ideal case $\phi \left( \omega _{0}\right) =-3\pi /2$
because $d\phi \left( \omega \right) /d\omega $ is very weak.

\section{Conclusion}
A variational method based on a coarse-grained operation has been used to
investigate in details the stability of an oscillating tip-cantilever system
near a surface. The tip-surface interaction is described by Van der Waals
forces. Results show that the resonance peak of the oscillator can be
described from two branches. The first one, named $u_{+}$, corresponds to
frequencies larger than the resonance. Stability criterions deduced foresee
that it is always stable. The second one, $u_{-}$, may be decomposed into
three domains~: two are stable and one is unstable. The second stable domain
of $u_{-}$ is small and is defined at the upper extremity of the resonance
peak. The phase at the resonance $\varphi \left( \omega _{0}\right) =-\pi /2$
is at the overlap of the $u_{+}$ and of this former second stable domain of $%
u_{-}$, thus the setpoint $\varphi \left( \omega _{0}\right) =-\pi /2$
belongs to a stable zone.

This result is of great importance to understand the stability in NC-AFM. In
this technique, the phase of the cantilever is adjusted to $-\pi /2$ within
an electronic feedback loop as the tip-surface distance $D$ is infinite. In
the approach mode, the frequency of the loop decreases, consequently the
phase becomes smaller than $-\pi /2$ because the slope $d\phi \left( \omega
\right) /d\omega $ of the phase of the phase shifter transfer function is
always negative. Thus the oscillator always ``slides'' along $u_{+}$ and the
system is unconditionally stable. This is what is usually observed
experimentally and confirmed by the results of the virtual NC-AFM we have
built. Because the slope $d\phi \left( \omega \right) /d\omega $ and the
frequency shift are very weak, we may consider that the phase $\varphi
\left( \omega _{0}\right) $ of the oscillator is always very close to $-\pi
/2$, typical variations being less than $2.10^{-3}$~rd. Consequently, the
damping signal keeps constant if no dissipative force is introduced in the
tip-surface interaction.

\section*{Referencecs}
\bibliographystyle{unsrt}

\bibliographystyle{NATURE}
\bibliography{amoi}

\section*{Figures}

\end{document}